\documentclass[useAMS,usegraphicx,usenatbib]{mn2e}

\usepackage{multirow}
\usepackage{amssymb}

\usepackage{amsmath}

\newcommand{\kms}{\mbox{$\>{\rm km\, s^{-1}}$}}
\newcommand{\pc}{\>{\rm pc}}
\newcommand{\kpc}{\mbox{$\>{\rm kpc}$}}

\newcommand{\Rd}{\mbox{$R_{\rm d}$}} 
\newcommand{\zd}{\mbox{$z_{\rm d}$}} 
\newcommand{\Md}{\mbox{$M_{\rm d}$}} 
\newcommand{\Mb}{\mbox{$M_{\rm b}$}} 
\newcommand\degrees{^\circ}
\newcommand{\avg}[1]{\mbox{$\left<{#1}\right>$}}

\newcommand{\lamost}{{\sc LAMOST}}
\newcommand{\rave}{{\sc RAVE}}
\newcommand{\segue}{{\sc SEGUE}}

\def\ie{{\it i.e.}}

\title[Vertical Structure and Kinematics of Spirals]{The Vertical
  Structure and Kinematics of Grand Design Spirals}
   
\author[Debattista]{Victor P. Debattista$^{1}$\thanks{E-mail:
    vpdebattista@gmail.com} \\ 
  $^1$ Jeremiah Horrocks Institute, University of Central Lancashire,
  Preston PR1 2HE, UK \\
}

\begin{document}   

\date{{\it Draft version on \today}}
\pagerange{\pageref{firstpage}--\pageref{lastpage}} \pubyear{----}
\maketitle

\label{firstpage}

\begin{abstract} 
  We use an $N$-body simulation to study the 3-D density distribution
  of spirals, and the resulting stellar vertical velocities.  Relative
  to the disc's rotation, the phase of the spiral's peak density away
  from the mid-plane trails that at the mid-plane.  In addition, at
  fixed radius the density distribution is azimuthally skewed, having
  a shallower slope on the trailing side inside corotation and
  switching to shallower on the leading side beyond corotation.
  The spirals induce non-zero average vertical velocities, \avg{V_z},
  as large as $\avg{V_z} \sim 10-20$ \kms, consistent with recent
  observations in the Milky Way.  The vertical motions are compressive
  (towards the mid-plane) as stars enter the spiral, and expanding
  (away from the mid-plane) as they leave it.  Since stars enter the
  spiral on the leading side outside corotation and on the trailing
  side within corotation, the relative phase of the expanding and
  compressive motions switches sides at corotation.  Moreover,
  because stars always enter the spiral on the shallow density
  gradient side and exit on the steeper side, the expanding motions
  are larger than the compressing motions.
\end{abstract}


%
%%%%%%%%%%%%%%%%%%%%%%%%%%%%%%%%%%%%%%%%%%%%%%%%%%%%%%%%%%%%%%%%%%%%%%%%%%%%%

\section{Introduction}
\label{sec:intro}

Spiral structure is one of the defining characteristics of disc
galaxies.  As a result, significant effort has been expended on
understanding the generation and dynamics of spirals.  The standard
interpretation views spirals as density waves in the stellar
distribution, but the cause of these waves, when not triggered by bars
or external perturbations, remains uncertain.  Models proposed include
the swing amplification of noise \citep{goldreich_lyndenbell65,
  julian_toomre66}, groove modes \citep{sellwood_lin89,
  sellwood_kahn91, sellwood12} or other modes \citep{lin_shu64},
possibly recurrent \citep{sellwood_carlberg14}.  Gas plays an
important role in the life cycle of spirals, both by cooling the disc,
allowing further generations of spirals \citep{sellwood_carlberg84},
and by providing granularity in the disc potential helpful in exciting
spirals \citep{donghia+13}.  The realization that spirals can drive
substantial radial mixing without heating \citep{sellwood_binney02,
  roskar+08a, schoenrich_binney09a} has led to a significant
resurgence of interest in the problem of spiral structure formation
and evolution.

\citet{widrow+12} reported an asymmetry in the density distribution
across the Milky Way's mid-plane from \segue\ data, which was explored
further by \citet{yanny_gardner13}.  Additionally, large-scale,
coherent vertical motions have been found in the Milky Way's disc in
\segue\ \citep{widrow+12}, \rave\ \citep{williams+13} and \lamost\
\citep{carlin+13} data.  Prior to these studies, \citet{bochanski+11}
had found a variation by spectral type in the vertical velocity of a
sample of M-dwarfs, although they could not distinguish these from the
effect of distance variations arising from differences in absolute
magnitude due to metallicity differences.  \citet{widrow+12} and
\citet{gomez+13} proposed that these asymmetries result from bending
caused by an interaction with substructure.  However, a puzzling
feature of the vertical velocities is that the dominant motion
corresponds to a breathing mode of the disc, \ie\ stars on either side
of the mid-plane are coherently moving either away from or towards the
mid-plane.  A bend in the disc caused by a perturber would instead
lead to motions that have the same direction on both sides of the
mid-plane.  \citet{schoenrich+12} cautioned that these motions may be
due to residual errors in the survey pipeline.  Nevertheless, in the
anti-centre direction, \citet{carlin+13} now find stellar motions
towards the mid-plane extending to 2 \kpc\ from the Sun.

These observations raise the question of whether internal causes are
possible and impel us to study the vertical structure and kinematics
of spirals.  To a large extent spirals have been studied in the
two-dimensional approximation under the simplifying assumption that
the motions of stars in the planar direction are decoupled from the
vertical motion.  Recently, \citet{faure+14} presented 3D test
particle orbits in a spiral potential, showing that spirals induce
vertical motions consistent with the Milky Way observations.  In this
Letter, we study the 3D density distribution of spirals and the effect
this has on the vertical motions using self-consistent $N$-body
simulations.

%%%%%%%%%%%%%%%%%%%%%%%%%%%%%%%%%%%%%%%%%%%%%%%%%%%%%%%%%%%%%%%%%%%%%%%%%%%%%

\section{Model Setup}
\label{sec:numerics}

We use simulation II of \citet{meidt+08} which was designed to have a
single grand design spiral with a well-defined pattern speed.  This
consists of a compact bulge and an exponential disc, immersed in a
spherical halo potential.  The disc has an exponential surface density
profile with mass \Md\ and scalelength \Rd; it is truncated at 5\Rd.
The vertical profile is Gaussian with a scaleheight $\zd = 0.1\Rd$.
We set the $Q$-parameter of the disc \citep{toomre81} to $Q=1.2$ using
the epicyclic approximation.  In order to slow the formation of a bar
till well after a spiral has formed, we include a massive compact
bulge.  The bulge was generated from an isotropic distribution
function (DF) of polytrope form $F(E) \propto (-E)^n$, with
$n=\frac{7}{2}$, using the method of \citet{prendergast_tomer70} as
described in \citet{debattista_sellwood00}.  The DF was integrated
iteratively until convergence in the global potential.  The bulge has
a mass $\Mb = \frac{1}{3}\Md$ and is truncated at $r_{\mathrm t} =
1.51 \Rd$.  The dark matter halo is represented by a rigid potential
of the form $\Phi_{\mathrm h} = \frac{1}{2}V_{\mathrm h}^2\ln(r^2 +
r_{\mathrm h}^2)$; we set $V_{\mathrm h} = 0.65$ and $r_{\mathrm h} =
5 \Rd$.

The disc is initially populated by 3 million particles.  We set up
particles in groups of four: the first particle in each quartet has
$(x,y,z,v_x,v_y,v_z)$ while the rest have $(-x,-y,z,-v_x,-v_y,v_z)$,
$(x,y,-z,v_x,v_y,-v_z)$ and $(-x,-y,-z,-v_x,-v_y,-v_z)$.  Besides
ensuring that the centre of the system does not move because of noise,
this quiet start \citep{sellwood83} has the desirable property
that the disc is highly symmetric about the mid-plane, ensuring that
no small-scale bends are present.  The main disadvantage is that it
reduces the number of independent particles, so that statistically it
is equivalent to only $\sim 7.5\times 10^5$ particles.  In order to
produce a strong spiral, we seed the disc with a groove mode
\citep{sellwood_lin89, sellwood_kahn91} by removing all particles in
the specific angular momentum range $1.6 < l_z < 1.8$ in units where
$\Rd = \Md = G = 1$.  This results in the removal of $5.6\%$ of the
disc's mass.

The system was evolved with the cylindrical polar grid code of
\citet{sellwood_valluri97}.  This solves for the potential by
expansion in a Fourier series in the $\phi$ direction, via a fast
Fourier transform in the vertical direction, and by convolution with
the Green function in the radial direction.  We use Fourier terms up
to $m=8$ in the potential solver (excluding $m=1$ in order that the
system remains centred at the origin).  The grid has $n_R \times
n_\phi \times n_z = 60 \times 64 \times 243$ grid cells.  The vertical
spacing of the grid cells is $0.125 \zd$, and the radial spacing is
logarithmic, reaching to $10 \Rd$.  Our timestep is $\delta t =
0.001$.  We set the softening length of all particles to $\epsilon =
0.017\Rd$.

We adopt a scaling to real units for facilitating comparison with the
Milky Way which has $\Rd =2.5\kpc$ and $V_{\mathrm h} = 228\kms$,
which produces a mean streaming velocity, \avg{V_\phi}, at 8 \kpc\ of
$\sim 220$ \kms.  This scaling corresponds to a unit of time equal to
7.1 Myr.  \citet{meidt+08} showed that the spiral that forms has a
corotation (CR) radius at $\sim 5 \kpc$ in our rescaled units.

%%%%%%%%%%%%%%%%%%%%%%%%%%%%%%%%%%%%%%%%%%%%%%%%%%%%%%%%%%%%%%%%%%%%%%%%%%%%%

\section{Results}
\label{sec:results}

\begin{figure}
%\centerline{
\includegraphics[angle=0.,width=\hsize]{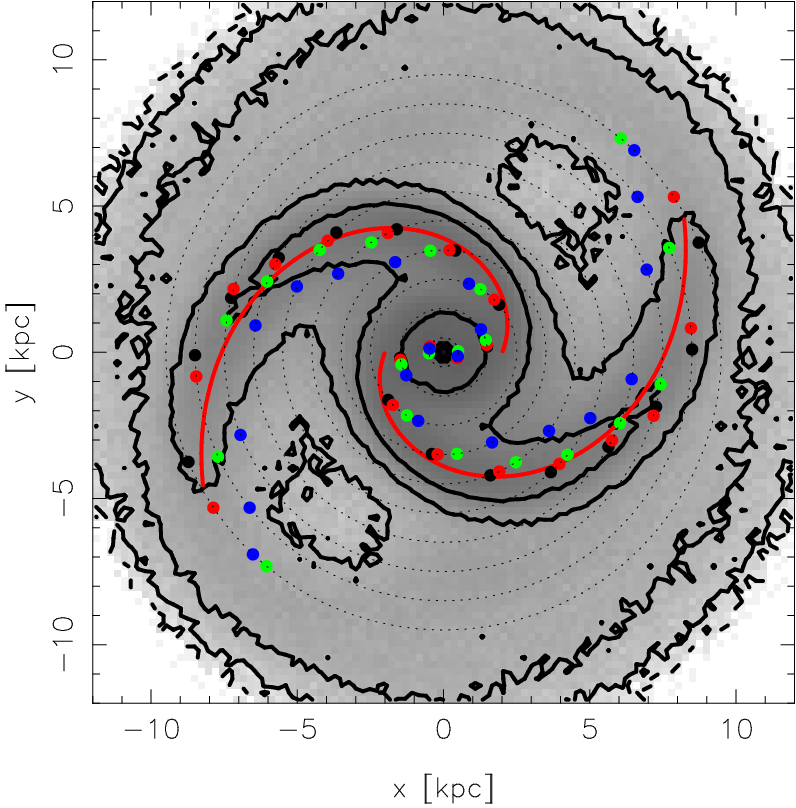} \\
\includegraphics[angle=-90.,width=\hsize]{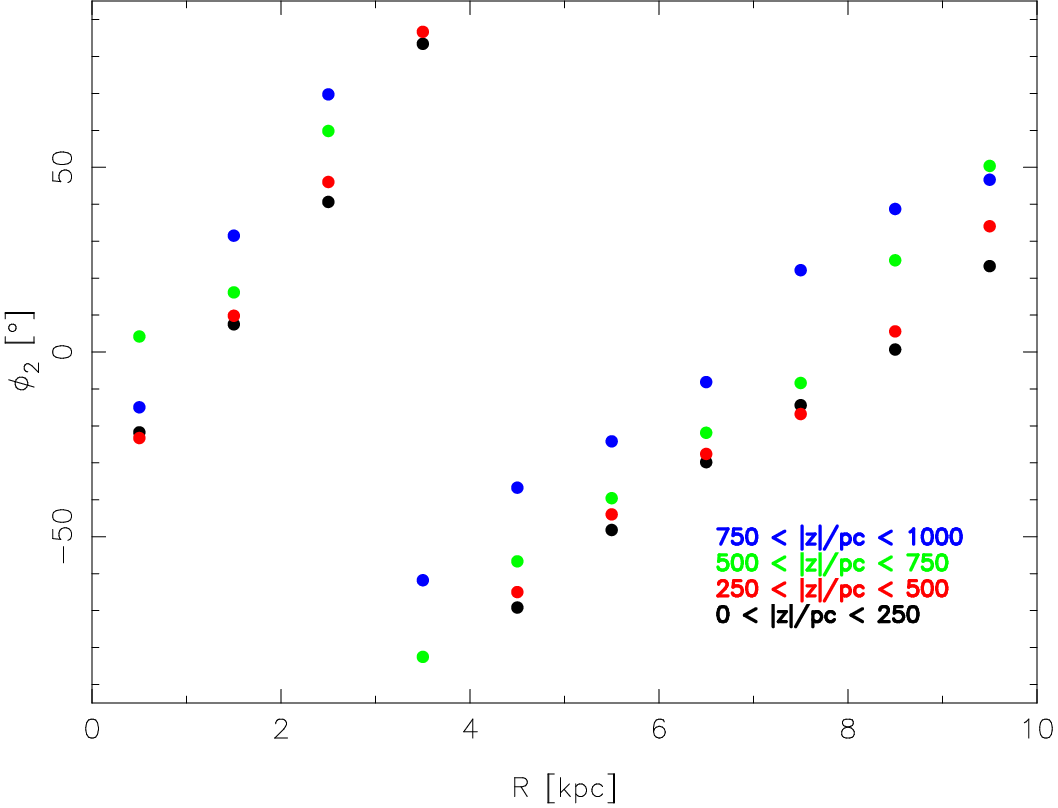}
%}
\caption{Top: surface density of the model at 3.8 Gyr.  The red lines
  show an $m=2$ logarithmic spiral of pitch angle $\gamma =
  40\degrees$.  The dotted circles represent the constant radius from
  0.5 to 9.5 kpc, in 1 kpc intervals. The filled circles indicate
  $\phi_2$ at that radius with $|z|$ colour-coded as in the bottom
  panel.  Bottom: phase of the $m=2$ perturbation for four slices in
  $|z|$ each of width $\delta z = \zd = 250$ pc.  Different slices are
  colour-coded as indicated.}
  \label{fig:density}
\end{figure}

Fig.~\ref{fig:density} shows the projected surface density after a
strong $m=2$ grand design spiral forms.  The spiral has maximum
surface density contrast $\delta \Sigma/\avg{\Sigma} \simeq 2.5$, or
peak Fourier $m=2$ amplitude $A_{\mathrm 2,peak} \simeq 0.65$ (see
fig.  13 of \cite{meidt+08}), making this a very strong, grand design
spiral \citep{rix_zaritsky95}.  Fig.~\ref{fig:density} overplots a
logarithmic spiral, $\phi \propto 2 \cot \gamma \ln R$, with pitch
angle $\gamma = 40\degrees$ which matches the peak surface density
well.  In comparison, the Milky Way's spirals have a pitch angle
$\gamma \sim 10\degrees$ \citep[e.g.][]{siebert+12}.

\subsection{Density distribution}

\begin{figure}
  \includegraphics[angle=0.,width=\hsize]{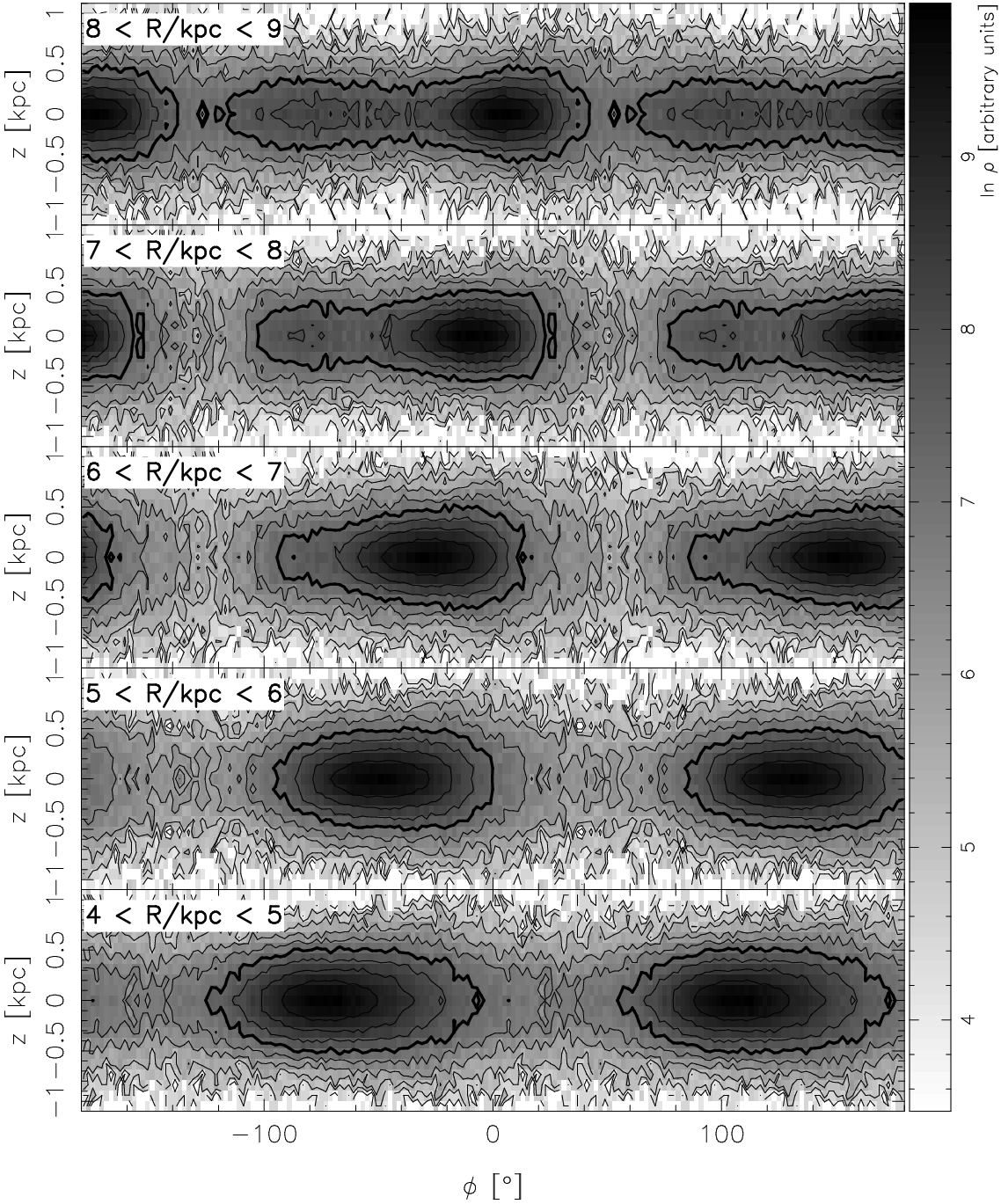}
  \caption{The stellar volume density in annuli as indicated in each
    panel.  The sense of rotation is towards decreasing $\phi$.
    Contours are separated by a factor of 1.9.  The bold contour
    represents a common value.}
\label{fig:phiz}
\end{figure}

Fig.~\ref{fig:phiz} maps the density in the $(\phi,z)$-plane for
1-\kpc-wide annuli.  At each radius, the peak density in the mid-plane
is ahead (relative to the direction of rotation) of the peak density
at higher $|z|$, resulting in wedge-shaped isodensity contours in
cross-section.  Since each annulus projects through a range of radii,
some of this offset may be caused by the radial superposition of a
winding spiral.  Therefore Fig.~\ref{fig:density} plots the phase of
the $m=2$ density moment, $\phi_2$, for slices in $|z|$ of width
$\delta z = \zd = 250$ pc.  The bottom panel shows these phases as a
function of radius, while the top panel plots these phases over the
surface density distribution.  Relative to the sense of rotation, the
phase lag from the mid-plane increases with height, from $\ga
5\degrees$ at $2-3\zd$ to $\ga 20\degrees$ by $3-4\zd$.

\begin{figure}
  \includegraphics[angle=-90.,width=\hsize]{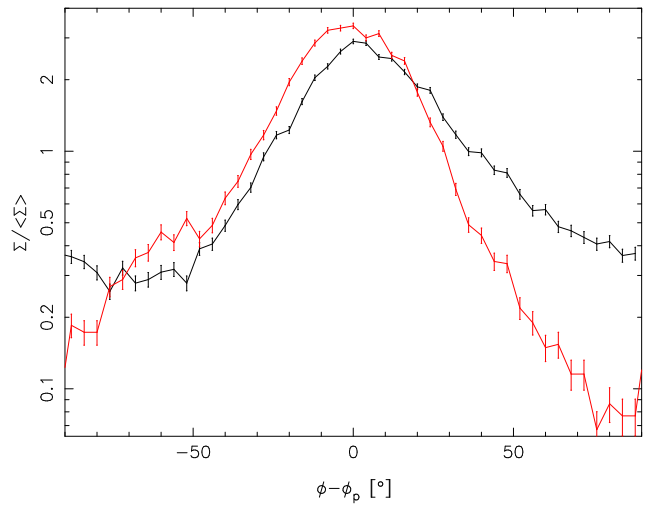}
  \caption{The density distribution as a function of phase relative to
    the phase of the peak density.  The black line shows the density
    distribution at 3.9 kpc while the red line shows the density at
    6.5 kpc.  In both cases, the density is computed for $|z| \leq 2
    \zd$.  Error bars indicate $\sqrt{N}$ uncertainty, where $N$ is
    the number of particles in each bin.}
\label{fig:skewness}
\end{figure}

Fig.~\ref{fig:phiz} also shows that the density distribution at fixed
$R$ and $z$ is asymmetric in $\phi$ relative to the peak density.
Fig.~\ref{fig:skewness} plots the density distribution as a function
of $\phi$ for two radii, one inside CR ($R=3.9$ kpc) and one outside
($R=6.5$ kpc), showing explicitly that the density distribution is
skewed.  The skewness changes sense across CR: inside, the shallower
side is behind the peak while outside it is ahead of the peak.  Thus,
stars always enter the spiral on the shallower density slope side and
exit it on the steeper side, regardless of which side of CR they are
on.  This asymmetry arises because as the density wave propagates
through the disc, stars on low eccentricity orbits are better able to
maintain phase with it, whereas stars on higher eccentricity orbits
trail the spiral.

\subsection{Vertical Kinematics}

\begin{figure*}
\centerline{
\includegraphics[angle=-90.,width=\hsize]{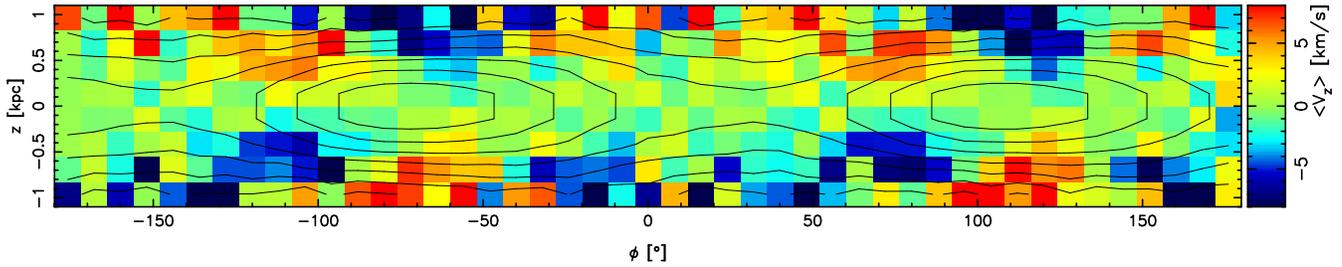}
}
\caption{Average vertical velocity, \avg{V_z}, in the annulus $4 \leq
  R/\kpc \leq 5$, just inside CR.  The sense of rotation is towards
  decreasing $\phi$. As in Fig. \ref{fig:phiz}, the volume density is
  indicated by the contours, which are separated by a factor of 2.3.}
\label{fig:phizkinematics}
\end{figure*}

We now turn to the trace of the spiral in the vertical velocities.
Fig. \ref{fig:phizkinematics} presents the mean vertical velocity,
\avg{V_z} in the annulus $4 \leq R/\kpc \leq 5$.  Generally, \avg{V_z}
is quite small, but non-zero.  Fig.  \ref{fig:phizkinematics} also
shows that the peak $|\avg{V_z}|$ increases with height.  The data of
\citet{widrow+12} and \citet{carlin+13} also show a trend of
increasing \avg{V_z} with $|z|$.

\begin{figure}
\centerline{
\includegraphics[angle=-90.,width=\hsize]{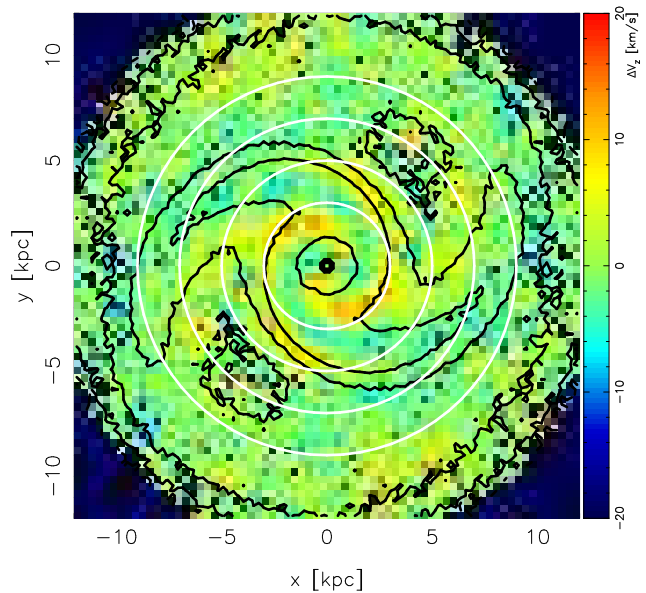} 
}
\caption{The asymmetry in $V_z$, $\Delta V_z$, defined in equation
  (\ref{eqn:velasym}), for $0 \leq |z| \leq 3\zd$.  The white circles
  range from 3 to 9 kpc in steps of 2 kpc, while the surface density
  is indicated by the solid black contours.  }
\label{fig:velasym}
\end{figure}

Fig.~\ref{fig:velasym} plots the vertical velocity asymmetry, which we
define as:
\begin{equation}
  \Delta V_z = \frac{\sum_\mathrm{bins,~ z>0}\avg{V_z} -
    \sum_\mathrm{bins,~ z<0}\avg{V_z}}{N_\mathrm{bins}/2}.
\label{eqn:velasym}
\end{equation}
Note that Equation \ref{eqn:velasym} first computes the average $V_z$
in each vertical bin before calculating the difference across the
mid-plane; as a result, $\Delta V_z$ is not dominated by the
high-density bins nearest $z=0$.  Wherever $\Delta V_z$ is negative
the motions are compressive, whereas they are expanding when it is
positive.  We use six bins in $z$, in the range $-3\zd \leq z \leq
3\zd$.  Non-zero vertical motions can clearly be seen in
Fig.~\ref{fig:velasym}.  The relative sense of compression versus
expansion changes across the CR radius; inside this radius vertical
motions are compressive behind the spiral peak and expanding ahead of
it.  Outside CR the sense shifts and compression (expansion) happens
ahead (behind) the spiral peak.

\begin{figure*}
\includegraphics[angle=0.,width=\hsize]{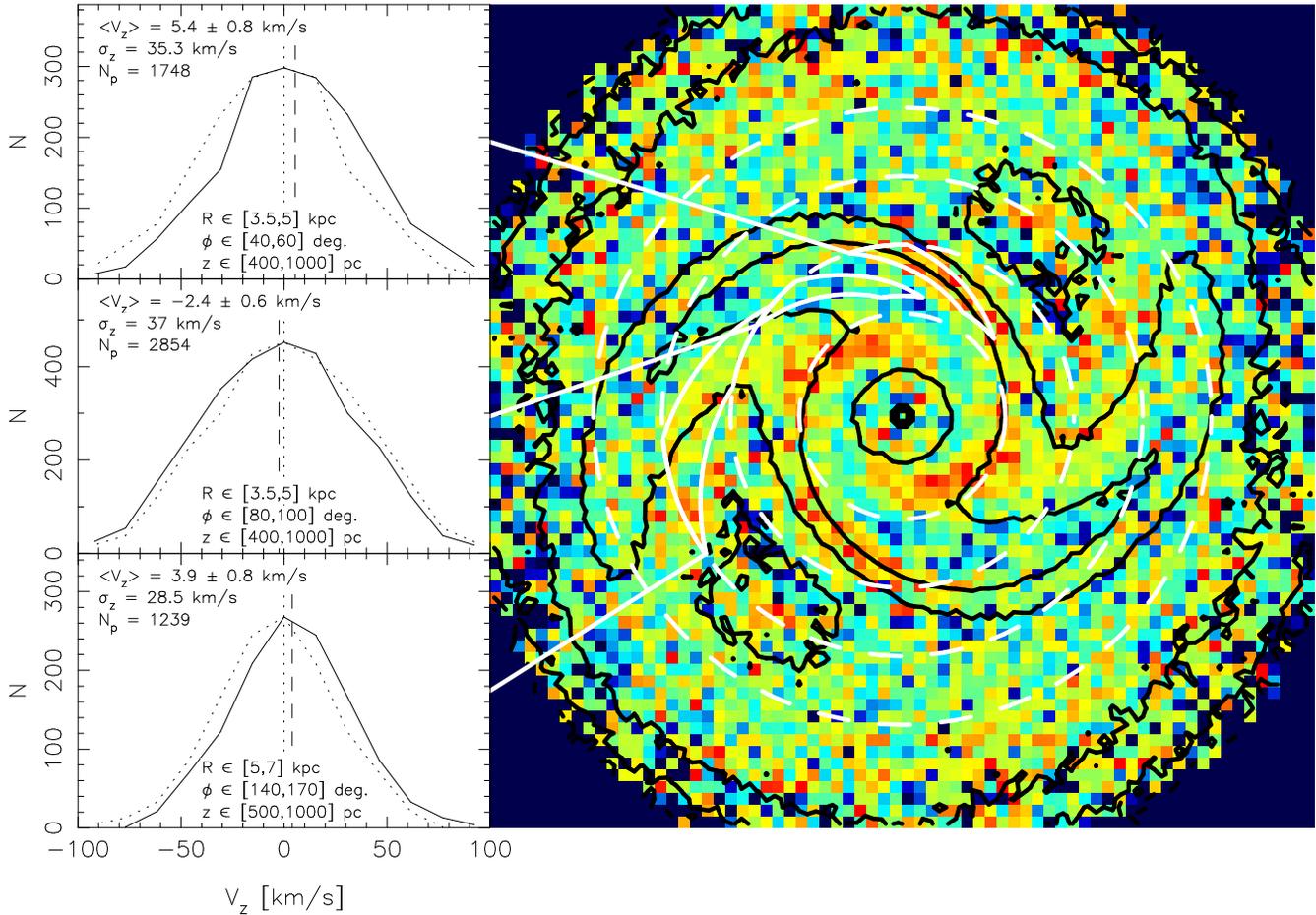}
\caption{Right: the map of $\Delta V_z$ for $\zd \leq |z| \leq 3\zd$,
  with the same colour scheme as in Fig. \ref{fig:velasym}.  The three
  left-hand panels show the distributions of $V_z$ (solid curves)
  selected from three regions following a log-spiral curve, and
  delineated by the solid white lines in the right-hand panel.  Each
  panel lists the number of star particles in each bin, $N_{\mathrm
    p}$, and the average vertical velocity, \avg{V_z}\ (indicated by
  the vertical dashed line), together with its uncertainty and the
  vertical velocity dispersion, $\sigma_z$.  The vertical dotted line
  shows $V_z = 0$, while the dotted curve shows the velocity
  distribution reflected about this line.}
\label{fig:losvds}
\end{figure*}

Because $|\avg{V_z}|$ is quite small close to the mid-plane (typically
$\sim 5 \kms$), in the right-hand panel of Fig.~\ref{fig:losvds} we
show $\Delta V_z$ excluding the two bins straddling the mid-plane
($|z| \leq \zd = 250 \pc$).  The velocity asymmetry now stands out
even more clearly and reaches values as large as 20 \kms.  The largest
expanding velocities are inside the CR radius on the leading edge of
the spiral, as can be seen in the annulus $3 \leq R/\kpc \leq 5$.
Since the stars overtake the spiral inside CR and are overtaken by it
outside CR, the expanding velocities are just ahead of the spiral
inside CR and behind it outside this radius.  This is most evident by
comparing $\Delta V_z$ in the two annuli at $3 \leq R/\kpc \leq 5$ and
$5 \leq R/\kpc \leq 7$.

The left-hand panels of Fig.~\ref{fig:losvds} show the distributions
of $V_z$ in three regions of the disc with large $|\Delta V_z|$.  In
all three cases, we select particles at some distance above the
mid-plane only.  Each of the bins has a non-symmetric distribution,
with \avg{V_z} that is statistically different from zero at more than
$3 \sigma$.  Reflecting these distributions about the $V_z = 0$ axis
leads to systematically offset distributions, indicating that the
non-zero averages are not due to noise.  The two bins at $3.5 \leq
R/\kpc \leq 5$ show the vertical kinematics inside CR on the
compressing and expanding sides of the spiral.  These two bins have
the same shape, but the $V_z$ distributions are different.  On the
expanding side $|\avg{V_z}| = 5.4 \pm 0.8 \kms$, while the compressing
side has less than half this value of $|\avg{V_z}|$.  This difference
arises because particles leaving the spiral always encounter a more
rapidly varying potential than when they enter the spiral.

%%%%%%%%%%%%%%%%%%%%%%%%%%%%%%%%%%%%%%%%%%%%%%%%%%%%%%%%%%%%%%%%%%%%%%%%%%%%%

\section{Relevance to the Milky Way}
\label{sec:discussion}

We have studied the 3D density distribution of strong, grand design
spirals using an $N$-body simulation.  At any given radius, the phase
of the peak density distribution varies with height, trailing that in
the mid-plane with increasing height.  At fixed height and radius, the
density variation in cylindrical angle $\phi$ is skewed.  Inside CR,
the density rises faster on the leading side of the spiral than on the
trailing.  This trend reverses outside CR.  These variations with
height and angle have not yet been observed in the Milky Way but may
be accessible to {\it Gaia} \citep{gaia} and the Large Synoptic Survey
Telescope \citep{lsst}.  The change in the sense of the skewness
across CR, independent of height above the mid-plane, permits
measurement of the Milky Way's spiral pattern speeds purely from the
density distribution.

The density variations in spirals lead to \avg{V_z}\ of the order of
$\sim 5-20 \kms$, increasing with $|z|$.  In the stationary frame of
the spiral, the motions are compressive as stars enter the spiral arm
and expanding as they exit.  The expanding motions therefore shift
from the leading to the trailing side of the spiral at the CR radius.
The recent test particle integrations of \citet{faure+14} find that
spirals induce vertical motions broadly consistent with those found
here, including the phase switch across CR.  Additionally, in the
corotating frame, stars enter the spiral on the shallow density
gradient side and exit on the steeper side.  As a result, the
expansion velocities tend to have larger amplitudes.  We also found
that \avg{V_z} increases with height above the mid-plane, in common
with the observations.  Although the spiral in our simulation is quite
strong and open, we found qualitatively similar behaviour in other
simulations.

The vertical motions we found are of similar amplitudes as those
observed in the Milky Way.  The observed vertical motions have been
interpreted as signs of bending waves in the disc, perhaps excited by
the Sagittarius dwarf \citep{gomez+13}.  Our highly symmetric disc
simulation expressly excludes such bending waves showing that spirals
provide an alternative explanation for at least the anti-symmetric
part of the observed kinematics.  Our simulation shows that these
motions are present also when the disc is self-gravitating and the
spirals are transient.  Weaker, more tightly wound spirals, such as
those in the Milky Way, may induce smaller motions but the qualitative
behaviour should remain similar.

\bigskip
\noindent
{\bf Acknowledgements.} 

\noindent
VPD is supported by STFC Consolidated grant no. ST/J001341/1.  This
project was started during a visit to the Aspen Center for Physics,
which is supported by the National Science Foundation under Grant No.
PHY-1066293.  We thank Rok Ro{\v s}kar and Jerry Sellwood for comments
on an earlier draft of the Letter and the anonymous referee for a
report that helped improve the clarity of the Letter.

\bigskip 
\noindent

\bibliographystyle{aj}
%\bibliography{ms.bbl}
\bibliography{allrefs}

\label{lastpage}

\end{document}